\documentclass[12pt,a4paper,leqno]{article}
\usepackage{graphicx}
\usepackage{amsmath}
\usepackage{amssymb}
\usepackage{amsthm}
\theoremstyle{plain}

\numberwithin{equation}{section}
\title{Modified method of simplest equation for obtaining 
exact solutions of nonlinear partial differential equations: past and present}
\author{Nikolay K. Vitanov}
\date{Institute of Mechanics, Bulgarian Academy of Sciences,
Akad. G. Bonchev Str., Bl. 4, 1113 Sofia, Bulgaria}
\begin{document}
\maketitle
\begin{abstract}
	We present a short review of the evolution of the methodology of the Method of simplest equation for obtaining exact particular
	solutions of nonlinear partial differential equations (NPDEs) and the recent extension
	of a version of this methodology called Modified method of simplest equation. This extension makes the methodology 
	capable to lead  to solutions of nonlinear partial differential equations that are more complicated than a single solitary
	wave. 	
\end{abstract}
\section{Introduction}
Differential equations occur frequently in the mathematical modeling of many  problems from natural and social sciences, e.g., 
fluid mechanics, plasma physics, atmospheric and ocean sciences, chemistry, materials science, biology, economics, social dynamics, 
etc. \cite{hirota} - \cite{vsd2}. Often the model equations are nonlinear partial differential equations and by means of 
the exact solutions of these equations one : (i) can understand complex nonlinear phenomena such as existence 
and change of different regimes of functioning of complex systems, transfer processes, etc. or (ii)  can test computer 
programs for numerical simulations by comparison of the obtained numerical results
to the corresponding exact solutions. Because of the above the exact solutions of NPDEs are studied very intensively 
\cite{ablowitz1} - \cite{tabor}. 
\par
The research on  the methodology for obtaining exact solutions of NPDEs started by
search for transformations that  transform the solved nonlinear partial differential equation to a
linear differential equation. The Hopf-Cole transformation \cite{hopf}, \cite{cole}  transforms the nonlinear Burgers 
equation to the linear heat equation and  the numerous 
attempts for obtaining such transformations led to the
\emph{Method of Inverse Scattering Transform} \cite{ablowitz1}, \cite{gardner}. 
Almost parallel to this
Hirota  developed a
direct method for obtaining of  exact solutions of NPDEs -  \emph{Hirota method} \cite{hirota}, \cite{hirota1} that
is based on bilinearization of the solved nonlinear partial differential equation by means of appropriate transformations. 
Truncated Painleve expansions may lead to many of these appropriate transformations \cite{tabor}, \cite{ct1} - \cite{wtk}. 
The line of research of interest for us below emerged when Kudryashov \cite{k3} studied the possibility
for obtaining exact solutions of NPDEs by a truncated Painleve expansion where the truncation happens after the "constant term" (i.e.,
the constant term is kept in the expansion).  Kudryashov  formulated   the \emph{Method of Simplest Equation (MSE)} \cite{k05} based 
on determination of singularity order $n$ of the solved NPDE and searching of
particular solution of this equation as series containing powers of solutions
of a simpler equation called \emph{simplest equation}. Kudryashov and Loguinova \cite{kl08}
extended the methodology and applied it for obtaining traveling wave solutions of  NPDEs. ( for
several examples see \cite{k5a} - \cite{k15}). 
\par
Below we shall review the results of our contribution to the method of simplest equation from our 
first research on the use of the ordinary differential 
equation of Bernoulli as simplest equation \cite{v10} and by application of the method to ecology
and population dynamics \cite{vd10} where we have used the concept of the balance equation to our last 
version of the method based on more than one simplest equations and eventually to more than one balance equation.
We note that the version called called \emph{Modified Method of Simplest Equation - MMSE} \cite{vdk}, \cite{v11} based on determination of the kind
of the simplest equation and truncation of the series of solutions of the simplest equation by means 
of application of a balance equation is equivalent of the \emph{Method of simplest equation}. 
Up to now our contributions to the methodology and its application are connected to the \emph{MMSE} \
\cite{v11a} - \cite{vdv17}. We note especially the article \cite{vdv15} where we have extended the methodology 
of the \emph{MMSE} to simplest equations of the class
	\begin{equation}\label{sf}
	\left (\frac{d^k g}{d\xi^k} \right)^l = \sum \limits_{j=0}^{m} d_j g^j
	\end{equation}
where $k=1,\dots$, $l =1,\dots$, and $m$ and $d_j$ are parameters. The solution of Eq.(\ref{sf}) defines
a special function that contains as particular cases, e.g.,: (i) trigonometric functions; (ii) hyperbolic functions;
(iii) elliptic functions of Jacobi; (iv) elliptic function of Weierstrass.
\section{MMSE and its version based on a single simplest equation - MMSE1 and on more than
one simplest equation - MMSEn}
\begin{figure}[htb!]
\centering
\includegraphics[scale=.5]{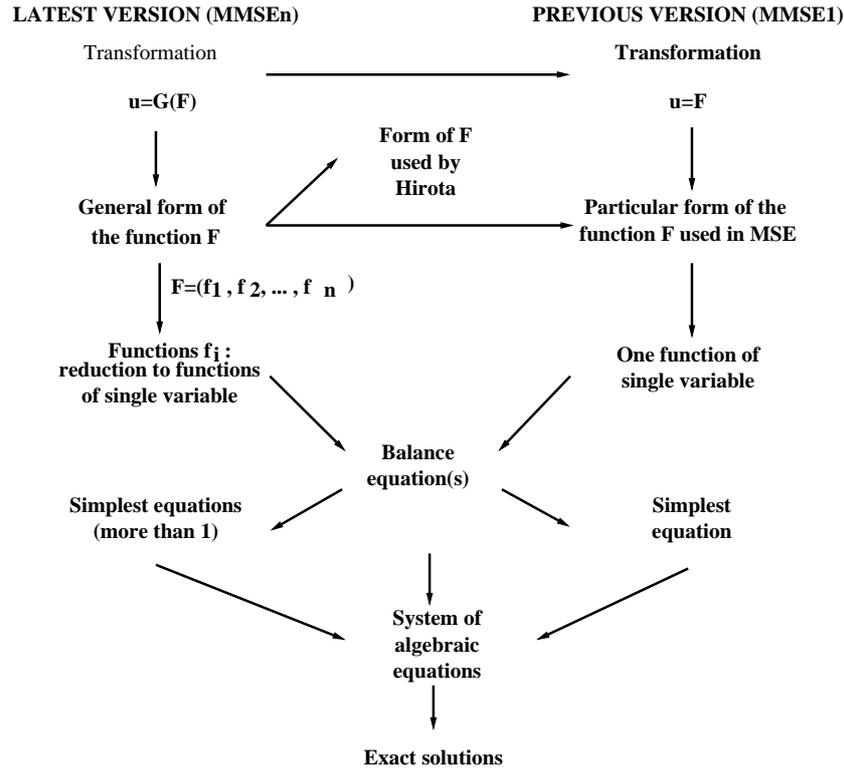}
\caption{Versions on the MMSE. MMSE1 is based on 1 simplest equation and 1 balance
equation. MMSEn is based on $n$ simplest equations ($n \ge 1$) and on one or more than
one balance equation. Note that in the two versions of the method the relationship
between the solution of the solved NPDE and the simplest equation(s) is different.}
\end{figure}
\par
The methodology of MMSE1 is as follows.
Let us consider a nonlinear partial differential equation 
\begin{equation}\label{eqx}
{\cal{N}}(u,\dots)=0
\end{equation}
where ${\cal{N}}(u,\dots)$ depends on the function $u(x,t)$
and some of its derivatives participate in  ($u$ can be a function of more than 1 spatial coordinate).
The steps of the methodology of the modified method of simplest equation for obtaining particular traveling wave 
solutions of a NPDE are:
\begin{enumerate}
	\item
	By means of the traveling wave ansatz $\xi = \alpha x + beta t$ ($\alpha$ and $\beta$ are parameters)
	$u(x,t)$ is transformed to $u(\xi)$ which is represented as a function of other function $f$
	that is solution of some ordinary differential equation (the 
	simplest equation). The form of the function $F(f)$ is can be different. One example is
	\begin{equation}\label{mx2}
	u = \sum \limits_{i=-M}^N \mu_n f^n 
	\end{equation}
	$\mu$ is a parameter. In the most cases one uses $M=0$.
	\item
	The application of Eq.(\ref{mx2}) to Eq.(\ref{eqx}) transforms the left-hand side of 
	this equation. Let the result of this transformation  be a function that is a sum of terms where each 
	term contains some function multiplied by a coefficient. This coefficient contains some of the 
	parameters of the solved equation and some of the parameters of the solution. In the most cases
	a balance procedure must be applied in order to ensure that the above-mentioned relationships
	for the coefficients contain more than one term ( e.g., if the result of the transformation 
	is a polynomial then the balance procedure has to ensure that the coefficient of each 
	term of the polynomial is a relationship that contains at least two terms).
	This balance procedure leads to one  relationship among the parameters 
	of the solved equation and parameters of the solution. The  relationship is called 
	\emph{balance equation}. 
	\item
	We may obtain a nontrivial solution of Eq. (\ref{eqx})  if all coefficients mentioned above are
	set to $0$. This condition usually leads to a system of nonlinear algebraic equations for the 
	coefficients of the solved nonlinear PDE and for the coefficients of the solution. Any nontrivial 
	solution of this algebraic system leads to a solution the studied  nonlinear partial differential 
	equation. Usually the above system of algebraic equations contains many equations that have to 
	be solved with the help of   a computer algebra system. 
\end{enumerate}
\par 
One can mention use of elements  of MMSE1 in some of our first publications \cite{mv1} - \cite{mv5}.
Our interest in the research on the exact solutions of nonlinear PDEs was revived
when we have encountered such an equation as model of traveling waves in population dynamics \cite{dop2}. 
We analyzed the equation
\begin{equation}\label{eq1}
\frac{\partial Q}{\partial t} - D \frac{\partial^2 Q}{\partial x^2} = EQ^4 + FQ^3 + GQ^2 + HQ
\end{equation}
where $D,E,F,G,H$ are parameters. In order to obtain the exact traveling wave solution of this equation we have used representation of the solution as finite series of power of solutions of Bernoulli equation and the concept of balance equation. The methodology used in
\cite{dop2} was used in \cite{dop1} for the case of traveling waves
in a system of two interacting populations. The next step was to search for exact solutions of more complicated  equations.
This was made for the class of equations \cite{v10}
\begin{equation}\label{eq2}
\sum \limits_{p=1}^{N_1} \alpha_p \frac{\partial ^p Q}{\partial t^p} +
\sum \limits_{q=1}^{N_2} \beta_q \frac{\partial^q Q}{\partial x^q} +
\sum \limits_{m=1}^M \mu_m Q^m =0
\end{equation}
where $\alpha_p$, $\beta_q$ and $\mu_m$ are parameters. Bernoulli and Riccati equations are used as simplest equations and the balance equation connected  to the class of equations (\ref{eq2}) is obtained.
The developed theory is used for obtaining exact solutions of equations of Ablowitz and Zepetella, Huxley equation for gene propagation, equation connected to biological invasions, and more complicated equations. 
\par
Further application of the methodology  was to the equations of the
reaction-diffusion class \cite{vd10}
\begin{equation}\label{eq3}
\frac{\partial Q}{\partial t} + \frac{dD}{dQ} \left( \frac{\partial Q}{\partial x} \right)^2 + D(Q) \frac{\partial^2 Q}{\partial x^2} + F(Q)=0
\end{equation}
and to reaction-telegraph class
\begin{equation}\label{eq4}
\frac{\partial Q}{\partial t} - \alpha \frac{\partial^2 Q}{\partial t^2} - \beta \frac{\partial^2 Q}{\partial x^2} -\gamma \frac{dF}{qQ}
\frac{\partial Q}{\partial t} - F(Q)=0
\end{equation}
Above $\alpha, \beta, \gamma$ are parameters and $D$ and $F$ depend on the population density $Q$. As simplest equations 
are used the equations of Bernoulli and Riccati. As first result here the corresponding balance equations for (\ref{eq3}) and (\ref{eq4}) are
obtained. Then numerous solutions are obtained for the two classes
of studied equations (\ref{eq3}) and (\ref{eq4}).
\par
Even more complicated class of equations was studied in \cite{vdk}, namely
\begin{eqnarray}\label{eq5}
\sum \limits_{p=1}^{N_1} A_p(Q) \frac{\partial^p Q}{\partial t^p} +
\sum \limits_{r=2}^{N_2} B_r(Q) \left( \frac{\partial Q}{\partial t}
\right)^r +
\sum \limits_{1=1}^{N_3} C_a(Q) \frac{\partial^s Q}{\partial x^s} +
\nonumber \\
\sum \limits_{u=2}^{N_4} D_u(Q) \left( \frac{\partial Q}{\partial x}
\right)^u + F(Q) = 0
\end{eqnarray}  
which contains as particular cases the reaction-diffusion and the reaction-telegraph equations.  As simplest equations
authors in \cite{vdk} use the equations of Bernoulli and Riccati. Any of these two simplest equations 
leads for corresponding balance equation for the class of equations (\ref{eq5}). 
\par
The capabilities of the MMSE1 have been demonstrated in \cite{v11} for the class of equations
\begin{eqnarray}\label{eq6}
\sum \limits_{i,j=0}^m A_{ij}(u) \frac{\partial^i u}{\partial t^i} \left(\frac{\partial u}{\partial t} \right)^j +
\sum \limits_{k,l=0}^n B_{ij}(u) \frac{\partial^k u}{\partial x^k} \left(\frac{\partial u}{\partial x} \right)^l = 0
\end{eqnarray}
Equations of Bernoulli and Riccati and their particular case (extended tahn - function equation) are used as simplest equations and corresponding balance equations are obtained. Exact solutions of two particular cases of Eq. (\ref{eq6}) (Swift-Hohenberg equation and 
generalized Rayleigh equation) are obtained.
\par
 The class of nonlinear PDEs that can be treated by the Modified method of simplest equation was extended in \cite{v11a}. 
 This class of equations is
 \begin{eqnarray}\label{eq7}
 \sum \limits_{i_1=0}^{\overline{n}_1}  \sum \limits_{i_2=0}^{{n}_1^*}
  \sum \limits_{j=0}^{n_2}  \sum \limits_{k_1=0}^{\overline{n}_3} \sum \limits_{k_2=0}^{n_3^*} \sum \limits_{l=0}^{n_4} \sum \limits_{p_1=0}^{\overline{n}_5} \sum \limits_{p_2=0}^{n_5^*}\sum \limits_{q=0}^{n_6} \sum \limits_{r_1=0}^{\overline{n}_7} \sum \limits_{r_2=0}^{n_7^*} \sum \limits_{s=0}^{n_8}
  \left(\frac{\partial^{i_1+i_2}u}{\partial x^{i_1} \partial t^{i_2}} \right)^j \cdot
  \left(\frac{\partial^{k_1+k_2}u}{\partial x^{k_1} \partial t^{k_2}} \right)^j \cdot \nonumber \\
   \left(\frac{\partial^{p_1+p_2}u}{\partial x^{p_1} \partial t^{p_2}} \right)^j \cdot
  \left(\frac{\partial^{r_1+r_2}u}{\partial x^{t_1} \partial t^{r_2}} \right)^j \cdot A_{i_1,i_2,j,k_1,k_2,l,p_1,p_2,q,r_1,r_2,s}(u) = G(u)
 \end{eqnarray}
where it was assumed that
$$
\frac{\partial^0 u}{\partial x^0} = \frac{\partial^0 u}{\partial t^0} =
\frac{\partial^0 u}{\partial x^0 \partial t^0} =0
$$
and G(u) and A(u) are polynomials
\begin{eqnarray}\label{eq8}
G(u) &=& \sum \limits_{\epsilon=0}^{\kappa} g_\epsilon u^\epsilon;
\nonumber \\
 A_{i_1,i_2,j,k_1,k_2,l,p_1,p_2,q,r_1,r_2,s}(u) &=& \sum \limits_{m=0}^{h_{i_1,i_2,j,k_1,k_2,l,p_1,p_2,q,r_1,r_2,s,m}}
a_{i_1,i_2,j,k_1,k_2,l,p_1,p_2,q,r_1,r_2,s,m} u^m \nonumber \\
\end{eqnarray}
The MMSE1 was applied to Eqs.(\ref{eq7})
and (\ref{eq8}) and the balance equations are obtained for the case when the solution is searched as power series constructed by solutions of a simplest equation. As illustration of the methodology exact solutions of the $b$-equations and of the generalized Degasperis - Processi equations are obtained.
\par 
The role of the simplest equation in the methodology of the MMSE1 is discussed in \cite{v11b}. As examples of simplest equations are discussed the equations of Bernoulli, Riccati and the differential equation for the elliptic functions of Jacobi. It is shown how the choice of the simplest equation influences the balance equation as well as the system of algebraic equations that is obtained by the application of the methodology of the modified method of simplest equation. It is shown that any nontrivial solution of 
certain system of nonlinear algebraic equations leads to exact traveling wave solution of a nonlinear partial differential equation.
As examples for obtaining exact solutions on the basis of Riccati equation as simplest equation one discussed the nonlinear partial differential equations of Newell - Whitehead and FitzHugh - Nagumo.
The general algorithm for obtaining differential equations that have 
exact traveling wave solutions constructed as power series of solutions of selected simplest equation is presented. The case of
use of differential equations for the Jacobi elliptic functions as
simplest equations was discussed further in \cite{pliska1}. Special attention to the exact traveling wave solutions of the nonlinear equations that are models for nonlinear water waves is given in \cite{vdk13} where exact traveling wave solutions are obtained for the
extended Korteweg-de Vries equation
\begin{equation}\label{eq9}
2 \frac{\partial \eta}{\partial \tau} + 3 \eta \frac{\partial \eta}{\partial \xi} + \frac{1}{3} \delta^2 \frac{\partial^3 \eta}{\partial \xi^3} - \frac{3}{4} \epsilon \eta^2 \frac{\partial \eta }{\partial \xi} = - \frac{1}{12} \epsilon \delta^2 \left( 23 \frac{\partial \eta}{\partial \xi} + 10 \eta \frac{\partial^3 \eta}{\partial \xi^3} \right) 
\end{equation}
(where $\epsilon$ and $\delta$ are small parameters called amplitude parameter and shallowness parameter) and for the generalized Camassa-Holm equation
\begin{equation}\label{eq10}
\frac{\partial u}{\partial t} + p_1 \frac{\partial u}{\partial x}
 +\frac{p_4}{2} \frac{\partial}{\partial x}g(u) - p_2 \frac{\partial^3 u}{\partial^2x \partial t}  - 2 p_3 \frac{\partial u}{\partial x}
 \frac{\partial^2 u}{\partial x^2} - p_3 u \frac{\partial^3 u}{\partial x^3} =0
 \end{equation}
for the case when the equation of Bernoulli and Riccati are used for simplest equations and for several forms of the polynomial $g(u)$.
\par 
Several interesting theorems have been proved in connection with the application of the Modified method of simplest equation. In \cite{vd14} the following theorem was proved:
\par
\textbf{Theorem:}
\emph{
	Let $\cal{P}$ be a polynomial of the function $u(x,t)$ and its derivatives.
	$u(x,t)$  belongs to the 
	differentiability class $C^k$, where $k$ is the highest order of derivative
	participating in $\cal{P}$. $\cal{P}$ can contain some or all of the following 
	parts:
	\textbf{(A)}
	polynomial of $u$;
	\textbf{(B)}
	monomials that contain derivatives of $u$ with respect to $x$ and/or products of
	such derivatives. Each such monomial can be multiplied by a polynomial
	of $u$;
	\textbf{(C)}
	monomials that contain derivatives of $u$ with respect to $t$ and/or products of
	such derivatives. Each such monomial can be multiplied by a polynomial
	of $u$;
	\textbf{(D)}
	monomials that contain mixed derivatives of $u$ with respect to $x$ and $t$ and/or products of such derivatives. Each such monomial can be multiplied by a polynomial
	of $u$;
	\textbf{(E)}
	monomials that contain products of derivatives of $u$ with respect to $x$ 
	and  derivatives of $u$ with respect to $t$. Each such monomial can be multiplied by a polynomial of $u$;
	\textbf{(F)}
	monomials that contain products of derivatives of $u$ with respect to $x$ 
	and mixed derivatives of $u$ with respect to $x$ and $t$. 
	Each such monomial can be multiplied by a polynomial of $u$;
	\textbf{(G)}
	monomials that contain products of derivatives of $u$ with respect to $t$ 
	and mixed derivatives of $u$ with respect to $x$ and $t$. 
	Each such monomial can be multiplied by a polynomial of $u$;
	\textbf{(H)}
	monomials that contain products of derivatives of $u$ with respect to 
	$x$, derivatives of $u$ with respect to $t$ and mixed derivatives of 
	$u$ with respect to $x$ and $t$. Each such monomial can be multiplied by 
	a polynomial of $u$.}
\par
\emph{
	Let us consider the nonlinear partial differential equation 
	\begin{equation}\label{tb1}
	{\cal{P}}=0.
	\end{equation}
	We search for solutions of this equation of the kind 
	$u(\xi)=u(\alpha x + \beta t) = u(\xi) = \gamma f(\xi)$, 
	where $\gamma$ is a parameter and $f(\xi)$ is solution of the simplest equation
	$f_{\xi}^2 = 4 (f^2 - f^3)$. The substitution of this solution in Eq.(\ref{tb1})
	leads to a relationship $\cal{R}$  of the kind
	\begin{equation}\label{tb5}
	{\cal{R}}= \sum_{i=0}^N C_i f(\xi)^i +f_{\xi} \left(\sum_{j=0}^M D_j f(\xi)^j \right)
	\end{equation}
	where $N$ and $M$ are natural numbers depending on the form of the polynomial $\cal{P}$ .
	The coefficients $C_i$ and $D_j$ depend on the parameters of Eq.(\ref{tb1}) and on $\alpha, \beta$
	and $\gamma$. Then each nontrivial solution of the nonlinear algebraic system
	\begin{equation}\label{tb7}
	C_i = 0, i=1,\dots,N; \ \ \ D_j = 0, j=1,\dots,M
	\end{equation}
	leads to solitary wave solution of the nonlinear partial differential equation (\ref{tb1}).
}
\par
We note hat the simplest equation from the above theorem, namely 	$f_{\xi}^2 = 4 (f^2 - f^3)$ has the solution $f(\xi)=\frac{1}{\cosh^2(\xi)}$ where $\xi = \alpha x + \beta t$. In other words the theorem gives us an information about the solitary wave solutions of large class of nonlinear partial differential equations.
\par 
An extension of the methodology with respect to more general simplest equation was made in \cite{vdv15}. There was proposed the use of the
simplest equation with polynomial nonlinearity (\ref{sf}). 
For the particular case $k=2$ in (\ref{sf}) and for a large class of nonlinear PDE with polynomial nonlinearity the Modified method of simplest equation is formulated in terms of calculation of sequences of polynomials. The methodology is applied to the generalized Kortweg-de Vries equation and to second order Korteweg - de Vries equation (called also Olver equation).
\par
In \cite{vdv17} the research presented in \cite{vd14} was extended for the case of simplest equation
$f_{\xi}^2 = n^2 (f^2 - f^(2n+2)/n)$ which has the solution $f(\xi)=\frac{1}{\cosh^n(\xi)}$ where $\xi = \alpha x + \beta t$.
\par 
The methodology of the modified method of simplest equation based on one simplest equation was applied in the last years for studying
propagation of waves in artery with aneurism \cite{e17} - \cite{e18a}.
\par 
The last modification of the modified method of simplest equation 
is connected to the possibility of use of more than one simplest equation that was applied in \cite{vd18}. 
This modification (MMSEn - Fig. 1) is as follows.
Let us consider a nonlinear partial differential equation (\ref{eqx})
where ${\cal{N}}(u,\dots)$ depends on the function $u(x,...,t)$
and some of its derivatives participate in  ($u$ can be a function of more than 1 spatial coordinate).
The 7 steps of the methodology of the modified method of simplest equation are as follows.
\begin{description}
	\item[1.)]
	We perform a transformation
	\begin{equation}\label{m1}
	u(x,\dots,t)=G(F(x,\dots,t))
	\end{equation}
	where $G(F)$ is some function of another function  $F$. In general
	$F(x,\dots,t)$ is a function of the spatial variables as well as of the time. The transformation $G(F)$
	may be the Painleve expansion \cite{hirota}, \cite{kudr90}, \cite{k3} or another transformation, e.g., $u(x,t)=4 \tan^{-1}[F(x,t)]$ for the case of the 
	sine - Gordon equation or $u(x,t) = 4 \tanh^{-1}[F(x,t)]$ for the case of sh-Gordon (Poisson-Boltzmann 
	equation) (for applications of the last two transformations see, e.g. \cite{mv1} - \cite{mv5}).
	In many particular cases one may skip this step (then we have just $u(x,\dots,t)=F(x,\dots,t)$) 
	but in numerous cases the step is necessary
	for obtaining a solution of the studied nonlinear PDE. The application of Eq.(\ref{m1}) to 
	Eq.(\ref{eqx}) leads to a nonlinear PDE for the function $F(x,\dots,t)$.
	\item[2.)]
	The function $F(x,\dots,t)$ is represented as a function of other functions $f_1,\dots,f_N$
	that are  connected to solutions of some differential equations (these equations can be partial 
	or ordinary differential equations) that are more simple than Eq.(\ref{eqx}). We note that 
	the possible values of $N$ are $N=1,2,\dots$ (there may be infinite number of functions $f$ too).
	The forms of the function $F(f_1,\dots,f_N)$ can be different. One example is
	\begin{eqnarray}\label{m2}
	F &=& \alpha + \sum \limits_{i_1=1}^N \beta_{i_1} f_{i_1} + \sum \limits_{i_1=1}^N  \sum \limits_{i_2=1}^N 
	\gamma_{i_1,i_2} f_{i_1} f_{i_2} + \dots + \nonumber \\
	&&\sum \limits_{i_1=1}^N \dots \sum \limits_{i_N=1}^N \sigma_{i_1,\dots,i_N} f_{i_1} \dots f_{i_N}
	\end{eqnarray}
	where $\alpha,\beta_{i_1}, \gamma_{i_1,i_2}, \sigma_{i_1,\dots,i_N}\dots  $ are parameters.
	We shall use Eq.(\ref{m2}) below. Note that the relationship (\ref{m2}) contains as particular case the 
	relationship used by Hirota \cite{hirota}. The power series $\sum \limits_{i=0}^N \mu_n f^n$ (where
	$\mu$ is a parameter) used in the previous versions of the methodology of the modified method of simplest equation are a particular case of the relationship (\ref{m2}) too.
	\item[3.)] 
	In general the functions $f_1,\dots,f_N$ are solutions of partial differential equations.
	By means of appropriate ans{\"a}tze (e.g.,  traveling-wave ans{\"a}tze such as 
	$\xi = \hat{\alpha} x + \hat{\beta} t$; $\zeta =\hat{\gamma} x + \hat{\delta} t$, 
	$\eta = \hat{\mu} y + \hat{\nu}t \dots$) 
	the solved   differential equations for $f_1,\dots,f_N$ may be reduced to   differential equations 
	$E_l$, containing derivatives of one or several functions
	\begin{equation}\label{i1}
	E_l \left[ a(\xi), a_{\xi},a_{\xi \xi},\dots, b(\zeta), b_\zeta, b_{\zeta \zeta}, \dots \right] = 0; \ \
	l=1,\dots,N
	\end{equation}
	In many cases (e.g, if the equations for the functions $f_1,\dots$ are ordinary differential equations) one may skip this step 
	but the step may be necessary if the equations for $f_1,\dots$ are partial differential equations.
	\item[4.)]
	We assume that	the functions $a(\xi)$, $b(\zeta)$, etc.,  are  functions of 
	other functions, e.g., $v(\xi)$, $w(\zeta)$, etc., i.e.
	\begin{equation}\label{i1x}
	a(\xi) = A[v(\xi)]; \ \ b(\zeta) = B[w(\zeta)]; \dots
	\end{equation} 
	Note that the kinds of the functions $A$ , $B$, $\dots$ are not prescribed. 
	Often one uses a finite-series relationship, e.g., 
	\begin{equation}\label{i2}
	a(\xi) = \sum_{\mu_1=-\nu_1}^{\nu_2} q_{\mu_1} [v (\xi)]^{\mu_1}; \ \ \ 
	b(\zeta) = \sum_{\mu_2=-\nu_3}^{\nu_4} r_{\mu_2} [w (\zeta)]^{\mu_2}, \dots 
	\end{equation}
	where $q_{\mu_1}$, $r_{\mu_2}$, $\dots$ are coefficients.
	However other kinds of relationships may be used too. 
	\item[5.)]
	The functions  $v(\xi)$, $w(\zeta)$, $\dots$ 
	are solutions of simpler ordinary differential equations called \emph{simplest equations}. 
	For several years the methodology of the modified method of simplest equation was based 
	on use of one simplest equation. This version of the methodology allows for the use of more
	than one simplest equation.
	\item[6.)]
	The application of the steps 1.) - 5.) to Eq.(\ref{eqx}) transforms the left-hand side of 
	this equation. Let the result of this transformation  be a function that is a sum of terms where each 
	term contains some function multiplied by a coefficient. This coefficient contains some of the 
	parameters of the solved equation and some of the parameters of the solution. In the most cases
	a balance procedure must be applied in order to ensure that the above-mentioned relationships
	for the coefficients contain more than one term ( e.g., if the result of the transformation 
	is a polynomial then the balance procedure has to ensure that the coefficient of each 
	term of the polynomial is a relationship that contains at least two terms).
	This balance procedure may lead to one or more additional relationships among the parameters 
	of the solved equation and parameters of the solution. The last relationships are called 
	\emph{balance equations}. 
	\item[7.)]
	We may obtain a nontrivial solution of Eq. (\ref{eqx})  if all coefficients mentioned in Step 6.) are
	set to $0$. This condition usually leads to a system of nonlinear algebraic equations for the 
	coefficients of the solved nonlinear PDE and for the coefficients of the solution. Any nontrivial 
	solution of this algebraic system leads to a solution the studied  nonlinear partial differential 
	equation. Usually the above system of algebraic equations contains many equations that have to 
	be solved with the help of   a computer algebra system. 
\end{description}
The above steps of the methodology are generalization of what was used in \cite{vd18} to obtain
exact traveling wave solutions of the nonlinear Schr{\"o}dinger equation. We note Eq.(\ref{m2})
that represents the relationship among the solution of the solved NPFE and the solutions of
the corresponding simplest equations. This relationship from MMSEn contains as particular
case the corresponding relationship from MMSE1 as well as the relationship used in \cite{hirota}.
\section{Concluding remarks}
There exists also research on the relation between the MMSE1 and other methods fro obtaining
exact solutions of NPDEs. Results of this research can be found in \cite{d12} -\cite{d13}. 
This research shows that MMSE1 is powerful method for obtaining exact
particular solutions of many NPDEs. The extension to MMSEn makes the methodology
even more useful as it becomes capable to obtain solutions that are more complicated
in comparison to solitary waves (if such solutions do exist). The main characteristic of MMSEn
is the possibility of use of more than one simplest
equation. MMSEn includes also a possibility for a transformation connected to  the searched solution. 
In such a way the possibility for use of a Painleve expansion or other transformations  
is presented in the methodology of MMSEn. This possibility 
in combination with the possibility of use of more than one simplest equation adds
the capability for obtaining multisolitons by the discussed version of the 
methodology of the modified method of simplest equation. In addition we consider
the relationship (\ref{m2}) that is used to connect the solution of the solved 
nonlinear partial differential equation to solutions of more simple differential 
equations. The discussed version of the methodology allows for the use of more than one 
balance equation too. All above opens new horizons for application of MMSEn for obtaining
exact solutions of complicated nonlinear partial differential equations.


\begin{thebibliography}{99}
	
	\bibitem{hirota}
	\textsc{Hirota, R.} Exact solution of Korteweg-de Vries equation for
	multiple collisions of solitons. Phys. Rev. Lett., \textbf{27}, (1971),  1192 -- 1194.
	\bibitem{debn}
	\textsc{Debnath, L.}  Nonlinear  partial differential equations for scientists and engineers. Springer, New York, 2012.
	\bibitem{hirsch}
	\textsc{Hirsch, M., R. L. Devaney, S. Smale.} Differential equations, dynamical systems, and an introduction to chaos. Academic Press, New York, 2004.
	\bibitem{strauss}
	\textsc{Strauss, W. A.} Partial differential equations: An introduction. Wiley, New York, 1992.
	\bibitem{knvit}
	\textsc{Vitanov, N. K.} Science dynamics and research production. Indicators, indexes, statistical laws and mathematical models. Springer, Cham, 2016.
	\bibitem{dop2}
	\textsc{Vitanov, N. K., I. P. Jordanov, Z. I. Dimitrova.} On nonlinear population waves. Applied Mathematics and Computation, \textbf{215}, (2009), 2950 -- 2964. 
	\bibitem{dop1}
	\textsc{Vitanov, N. K. I. P. Jordanov, Z. I. Dimitrova.} On nonlinear dynamics of interacting populations: Coupled kink waves in a system of two populations. 
	Communications in Nonlinear Science and Numerical Simulation, \textbf{14}, (2009) , 2379 - 2388. 
	\bibitem{ma1}
	\textsc{Gerardy, J. M., M. Ausloos}. Absorption spectrum of clusters of spheres from the general solution of Maxwell's equations. II. Optical properties of aggregated metal spheres. Physical Review B,
	\textbf{25}, (1982), 4204 -- 4229.
	\bibitem{dop3}
	\textsc{Dimitrova, Z. I. N. K. Vitanov.} Adaptation and its impact on the dynamics of a system of three competing populations. Physica A,  \textbf{300}, (2001), 91 -- 115.
	\bibitem{ma2}
	\textsc{Caram, L. F., C. F. Caiafa, A. N. Proto,  M. Ausloos}. Dynamic peer-to-peer competition.
	Physica A, \textbf{389} (2010), 2628 -- 2636.
	\bibitem{dop4}
	\textsc{Dimitrova, Z. I. N. K. Vitanov.} Influence of adaptation on the nonlinear dynamics of a system of competing populations.
	Physics Letters A, \textbf{272}, (2000), 368 -- 380.
	\bibitem{ma3}
	\textsc{Ausloos, M., F. Petroni}. Statistical dynamics of religion evolutions. Physica A,
	\textbf{388}, (2009), 4438 -- 4444.
	\bibitem{dop5}
	\textsc{Vitanov, N. K. Z. I. Dimitrova, M. Ausloos.} Verhulst–Lotka–Volterra (VLV) model of ideological struggle.
	Physica A: Statistical Mechanics and its Applications, \textbf{389}, (2010), 4970 -- 4980.
	\bibitem{dop6}
	\textsc{Dimitrova, Z. I., N. K. Vitanov.} Chaotic pairwise competition.
	Theoretical Population Biology,  \textbf{66}, (2004), 1 -- 12.
	\bibitem{ma4}
	\textsc{Ausloos, M.} Another analytic view about quantifying social forces.  Advances in Complex Systems \textbf{16}, (2013) No.1, 1250088. 
	\bibitem{dop7}
	\textsc{Vitanov, N. K.,  F. H. Busse.} Bounds on the heat transport in a horizontal fluid layer with stress-free boundaries. Zeitschrift für angewandte Mathematik und Physik ZAMP, \textbf{48}, (1997), 310 -- 324.
	\bibitem{dop8}
	\textsc{Panchev, S.,   T. Spassova, N. K. Vitanov.} Analytical and numerical investigation of two families of Lorenz-like dynamical systems. Chaos, Solitons \& Fractals,  \textbf{33}, (2007), 1658 - 1671.
	\bibitem{dop9}
	\textsc{Vitanov, N. K., Z. I. Dimitrova, H. Kantz.} On the trap of extinction and its elimination. Physics Letters A, \textbf{349}, (2006), 350 - 355.
	\bibitem{vsd2}
	\textsc{Vitanov, N. K., K. N. Vitanov.} Box model of migration channels. Mathematical Social Sciences,  \textbf{80}, (2016), 108 -- 114.
	\bibitem{ablowitz1}
	\textsc{Ablowitz, M. J., D. J. Kaup, A. C. Newell, H. Segur.} Inverse scattering transform - Fourier analysis for nonlinear problems. Studies in Applied 
	Mathematics,  \textbf{53}, (1974) 249 -- 315 .	
	\bibitem{ac}
	\textsc{Ablowitz, M. J.,  P. A. Clarkson.} Solitons, nonlinear evolution equations and inverse scattering. Cambridge University Press, Cambridge, 1991.
	\bibitem{galakt}
	\textsc{Galaktionov, V. A., S. R. Svirhchevskii.} Exact solutions and invariant subspaces of nonlinear partial
	differential equations in mechanics and physics. Chapman \& Hall/CRC, Bora Raton, FL, 2007.
	\bibitem{gardner}
	\textsc{Gardner, C. S.,  J. M. Greene, M. D. Kruskal, R. R. Miura.} Method for solving Korteweg- de Vries equation. Phys. Rev. Lett., \textbf{19}, (1967) 
	1095 -- 1097.
	\bibitem{holmesx}
	\textsc{Holmes, P. J. L. Lumley, G. Berkooz.} Turbulence, coherent structures, dynamical systems and symmetry.  Cambridge University Press, Cambridge, 1996.
	\bibitem{kudr90}
	\textsc{Kudryashov, N. A.} Exact solutions of the generalized Kuramoto - Sivashinsky equation. Phys. Lett. A, \textbf{147}, (1990),  287 -- 291. 	
	\bibitem{logan}
	\textsc{Logan, J. D.} An introduction to nonlinear partial differential equations. Wiley, New York, 2008.	
	\bibitem{remos}
	\textsc{Remoissenet, M.} Waves called solitons. Springer, Berlin, 1993.
	\bibitem{tabor}
	\textsc{Tabor, M.} Chaos and integrability in dynamical systems.  Wiley, New York, 1989.
        \bibitem{hopf}
        \textsc{Hopf, E.} The partial differential equation $u_t + uu_x = \mu_{xx}$. Communications on Pure and Applied Mathematics, \textbf{3}, (1950) 201 -- 230.
	\bibitem{cole}
	\textsc{Cole, J. D.} On a quasi-linear parabolic equation occurring in aerodynamics. Quarterly of Applied 
	Mathematics, \textbf{9}, (1951)  225 -- 236	
	\bibitem{hirota1}
	\textsc{Hirota, R.} The direct method in soliton theory. Cambridge University Press, Cambridge, 2004.
	\bibitem{ct1}
	\textsc{Carrielo, F., M. Tabor.} Painleve expansions for nonlinear nonitegrable evolution equations. Physica D,  \textbf{39}, (1989) 77 -- 94.
	\bibitem{ct2}
	\textsc{Carrielo, F., M. Tabor.} Similarity reductions from extended Painleve expansion for nonitegrable evolution equations. Physica D, \textbf{53}, (1991) 59 -- 70.
	\bibitem{weiss1}
	\textsc{Weiss, J.} B{\"a}cklung transformations and the Painleve property. In R. Conte, N. Boccara (Eds.)
	Partially integrable evolution equations in physics, pp. 375 -- 411 , Kluwer, Dordrecht, 1990.
	\bibitem{wtk}
	\textsc{Weiss, J., M. Tabor, G. Carnevalle.} The Painleve property for partial differential equations.
	Journal of Mathematical Physics, \textbf{24}, (1983) 522 -- 526.
	\bibitem{k3}
	\textsc{Kudryashov, N. A.} On types of nonlinear nonitegrable equations with exact solutions. Physics Letters A, \textbf{155}, (1991)  269 -- 275.
	\bibitem{k05}
	\textsc{Kudryshov, N. A.} Simplest equation method to look for exact solutions of nonlinear differential 
	equations. Chaos, Solitons \& Fractals, \textbf{24}, (2005),  1217 -- 1231.
	\bibitem{kl08}
	\textsc{Kudryashov, N. A., N. B. Loguinova.} Extended simplest equation method for nonlinear differential  
	equations. Applied Mathematics and Computation, \textbf{205}, (2008) 361 -- 365.
	\bibitem{k5a}
	\textsc{Kudryashov, N. A.} Exact solitary waves of the Fisher equation. Physics Letters A, \textbf{342}, (2005)  99 -- 106.
	\bibitem{k8}
	\textsc{Kudryashov, N. A.} Solitary and periodic wave solutions of the generalized Kuramoto-Sivashinsky 
	equation. Regular and Chaotic Dynamics, \textbf{13}, (2008) 234 -- 238.
	\bibitem{k9}
	\textsc{Kudryashov, N. A., M. V. Demina.} Traveling wave solutions of the generalized nonlinear evolution 
	equations. Applied Mathematics and Computation, \textbf{201}, (2009)  551 -- 557.
	\bibitem{k12}
	\textsc{Kudryashov, N. A.} Exact solutions of the Swift-Hohenberg equation with dispersion.
	Communications in Nonlinear Science and Numerical Simulation, \textbf{17}, (2012) 
	26 -- 34.
	\bibitem{k12b}
	\textsc{Kudryashov, N. A.} One method for finding exact solutions of nonlinear differential equations.
	Communications in Nonlinear Science and Numerical Simulation, \textbf{17}, (2012)  
	2248 -- 2253.
	\bibitem{k12c}
	\textsc{Kudryashov, N. A.,  D. I. Sinelshchikov.} Nonlinear differential equations of the second, third and 
	fourth order with exact solutions. Applied Mathematics and Computation, \textbf{218},
	(2012),  10454 -- 10467.
	\bibitem{k15}
	\textsc{Kudryashov, N. A.} Logistic function as solution of many nonlinear differential equations.
	Applied Mathematical Modelling 39  (2015)  5733 -- 5742.
	\bibitem{v10}
	\textsc{Vitanov, N. K.} Application of simplest equations of Bernoulli and Riccati kind for obtaining exact
	traveling-wave solutions for a class of PDEs with polynomial nonlinearity. Communications in
	Nonlinear Science and Numerical Simulation 15 (2010)  2050 -- 2060.
	\bibitem{vd10}
	\textsc{Vitanov, N. K., Z. I. Dimitrova.} Application of the method of simplest equation for obtaining 
	exact traveling-wave solutions for two classes of model PDEs from ecology and population dynamics.
	Communications in Nonlinear Science and Numerical Simulation 15 (2010)   2836 -- 2845.
	\bibitem{vdk}
	\textsc{Vitanov, N. K., Z. I. Dimitrova, H. Kantz.} Modified method of simplest equation and its application 
	to nonlinear PDEs. Applied Mathematics and Computation, \textbf{216}, (2010) 
	2587 -- 2595
	\bibitem{v11}
	\textsc{Vitanov, N. K.} Modified method of simplest equation: powerful tool for obtaining exact and approximate 
	traveling-wave solutions of nonlinear partial differential equations. Communications in Nonlinear Science and Numerical Simulation, 
	\textbf{16}, (2011), 1176 -- 1185.
	\bibitem{v11a}
	\textsc{Vitanov, N. K., Z. I. Dimitrova, K. N. Vitanov.} On the class of nonlinear PDEs that can be treated by 
	the modified method of simplest equation. Application to generalized Degasperis–Processi equation 
	and b-equation. Communications in Nonlinear Science and Numerical Simulation, \textsc{16}, (2011),  3033 -- 3044.
	\bibitem{v11b}
	\textsc{Vitanov, N. K.} On modified method of simplest equation for obtaining exact and approximate 
	solutions of nonlinear PDEs: the role of the simplest equation. Communications in Nonlinear Science 
	and Numerical Simulation, \textsc{16}, (2011) 4215 -- 4231.
	\bibitem{pliska1}
	\textsc{Vitanov, N. K.} On modified method of simplest equation for obtaining exact solutions of nonlinear PDEs: case of elliptic simplest equation. Pliska Studia Mathematica Bulgarica \textbf{21}, (2012), 257 -- 266.
	\bibitem{vdk13}
	\textsc{Vitanov, N. K., Z. I. Dimitrova, H. Kantz.} Application of the method of simplest equation for 
	obtaining exact traveling-wave solutions for the extended Korteweg - de Vries equation and 
	generalized Camassa - Holm equation. Applied Mathematics and Computation, \textbf{219}, (2013),  7480 -- 7492.
	\bibitem{vdv13}
	\textsc{Vitanov, N. K., Z. I. Dimitrova, K. N. Vitanov.} Traveling waves and statistical distributions 
	connected to systems of interacting populations. Computers \& Mathematics with Applications, \textbf{66}, (2013)  1666 -- 1684.
	\bibitem{vd14}
	\textsc{Vitanov, N. K., Z. I. Dimitrova.} Solitary wave solutions for nonlinear partial differential 
	equations that contain monomials of odd and even grades with respect to participating 
	derivatives. Applied Mathematics and Computation, \textbf{247},  (2014)  213 -- 217.
	\bibitem{vdv17}
	\textsc{Vitanov, N. K., Z. I. Dimitrova, T. I. Ivanova.} On solitary wave solutions of a class of 
	nonlinear partial differential equations based on the function $1/cosh^n(\alpha x + \beta t)$.
	Applied Mathematics and Computation, \textbf{315}, (2017) 372 -- 380.
	\bibitem{vdv15}
	\textsc{Vitanov, N. K., Z. I. Dimitrova, K. N. Vitanov.} Modified method of simplest equation for 
	obtaining exact analytical solutions of nonlinear partial differential equations: Further 
	development of the methodology with applications. Applied Mathematics and Computation, \textbf{269}, (2015) 363 -- 378.
	\bibitem{mv1}
	\textsc{Martinov, N., N. Vitanov.} On the correspondence between the self-consistent 2D Poisson-Boltzmann structures and the sine-Gordon waves.
	Journal of Physics A: Mathematical and General, \textbf{25}, (1992),  L51 -- L56.
	\bibitem{mv2}
	\textsc{Martinov, N., N. Vitanov.} On some solutions of the two-dimensional sine-Gordon equation. Journal of Physics A: Mathematical and General, \textbf{25}, (1992), L419 -- L426.
	\bibitem{mv3}
	\textsc{Martinov, N. K., N. K. Vitanov.} New class of running-wave solutions of the (2+1)-dimensional sine-Gordon equation. Journal of Physics A: Mathematical and General, \textbf{27}, (1994), 4611 -- 4618.
	\bibitem{mv4}
	\textsc{Martinov, N. K., N. K. Vitanov.} On the self-consistent thermal equilibrium structures in two-dimensional negative-temperature systems. Canadian Journal of Physics, \textbf{72}, (1994), 618 -- 624.
	\bibitem{mv5}
	\textsc{Vitanov, N. K.} On travelling waves and double-periodic structures in two-dimensional sine-Gordon systems. Journal of Physics A: Mathematical and General, \textbf{29},  (1996),  5195 -- 5207.
	\bibitem{e17}
	\textsc{Nikolova, E. V., I. P. Jordanov, Z. I. Dimitrova, N. K. Vitanov.}
	Evolution of nonlinear waves in a blood-filled artery with an aneurysm. AIP Conference Proceedings, \textbf{1895}, (2017), 070002.
	\bibitem{e17a}
	\textsc{Nikolova, E. V.} Evolution equation for propagation of blood pressure waves in an artery with an aneurysm, pp. 327 - 339 in K. Georgiev, M. Todorov,
	I. Georgiev (Eds.) Advanced Computing in Industrial Mathematics.
	Springer, Cham, 2017.
	\bibitem{e18}
	\textsc{Nikolova, E. V., I. P. Jordanov, Z. I. Dimitrova, N. K. Vitanov.}
	Nonlinear evolution equation for propagation of waves in an artery with an aneurysm: An exact solution obtained by the modified method of simplest equation.  pp. 131 - 144 in K. Georgiev, M. Todorov,
	I. Georgiev (Eds.) Advanced Computing in Industrial Mathematics.
	Springer, Cham, 2018.
	\bibitem{e18a}
	\textsc{Nikolova, E.} On nonlinear waves in a blood-filled artery with an aneurysm. AIP Conference Proceedings, \textbf{1978}, (2018), 470050.
	\bibitem{vd18}
	\textsc{Vitanov, N. K.,Z. I. Dimitrova.} Modified method of simplest equation applied to the nonlinear Schrödinger equation. Journal of Theoretical and Applied Mechanics, Sofia, \textbf{48}, No. 1, (2018), 59  -- 68.
	\bibitem{d12}
	\textsc{Dimitrova, Z. I.} On traveling waves in lattices: The case of Riccati lattices.  Journal of Theoretical and Applied Mechanics, Sofia, \textbf{42}, No. 2, (2012),  3 -- 22.
	\bibitem{d12a}
	\textsc{Dimitrova, Z. I.} Relation between  G'/G- expansion method and the modified method of simplest equation. Comp. rend. Acad. bulg. Sci.
	\textbf{65}, No. 11, (2012), 1513 -- 1520.
	\bibitem{dv13}
	\textsc{Dimitrova, Z. I., K. N. Vitanov.} Integrability of differential equations with fluid mechanics application; from Painleve property to the method of simplest equation. Journal of Theoretical and Applied Mechanics, Sofia, \textbf{43}, No. 2, (2013),  31 -- 42. 
	\bibitem{d13}
	\textsc{Dimitrova, Z. I.} Discussion on exp-function method and modified method of simplest equation. Compt. rend. Acad. bulg. Sci. \textbf{66}, (2013), 975 -- 982.
	
\end{thebibliography}
\end{document}